\documentclass[10pt,amsmath,amssymb,nofootinbib,twoside,twocolumn,superscriptaddress,floats,floatfix,aps,preprintnumbers]{revtex4-2}
\usepackage[utf8]{inputenc}
\usepackage{babel} 
\usepackage{aas_macros}
\usepackage{amsmath} 
\usepackage{amsthm}
\usepackage{amssymb} 
\usepackage{physics}
\usepackage{braket} 
\usepackage{slashed} 
\usepackage{hyperref}
\usepackage{graphicx} 
\usepackage{mathrsfs} 
\usepackage{cancel} 
\usepackage{amsfonts} 
\usepackage{comment}
\usepackage{float} 
\usepackage{xcolor}

\begin{document}

\title{Formation and relaxation of halos in the context of wave DM particles evolving on a background of neutrino condensate}
\author{A. Capolupo}
\email{capolupo@sa.infn.it} \affiliation{Dipartimento di Fisica ``E.R. Caianiello'' Universit\'a di Salerno,
	and INFN -- Gruppo Collegato di Salerno, Via Giovanni Paolo II, 132,
	84084 Fisciano (SA), Italy}

\author{I. De Martino}
\email{ivan.demartino@usal.es}\affiliation{Universidad de Salamanca, Departamento de Física Fundamental, P. de la Merced, Salamanca, E-37008, España}
\affiliation{Insituto Universitario de F\'isica Fundamental y Matem\'aticas (IUFFyM), Universidad de Salamanca, Plaza de la Merced, s/n, E-37008 Salamanca, Spain}

\author{S. Monda}
\email{smonda@unisa.it}\affiliation{Dipartimento di Fisica ``E.R. Caianiello'' Universit\'a di Salerno,
and INFN -- Gruppo Collegato di Salerno, Via Giovanni Paolo II, 132,
84084 Fisciano (SA), Italy}

\author{R. Della Monica}
\email{rdellamonica@tecnico.ulisboa.pt}
\affiliation{CENTRA, Departamento de Física, Instituto Superior Técnico – IST\\
Universidade de Lisboa – UL, Avenida Rovisco Pais 1, 1049-001 Lisboa, Portugal}

\author{A. Quaranta}
\email{anquaranta@unisa.it}
\affiliation{University of Camerino, Via Madonna delle Carceri, Camerino, 62032, Italy.}

\affiliation{Dipartimento di Fisica ``E.R. Caianiello'' Universit\'a di Salerno,
and INFN -- Gruppo Collegato di Salerno, Via Giovanni Paolo II, 132,
84084 Fisciano (SA), Italy}

\begin{abstract}
We investigate the formation and relaxation of dark matter halos in the context of wave dark matter particles evolving on a background of neutrino condensate. To this aim, we
solved numerically the Schrödinger-Poisson system to model the dynamical evolution of ultralight bosonic dark matter particles in the presence of neutrino condensate. The latter appears as an additional source of the gravitational field in the Poisson equation, while its dynamical evolution and interaction with the environment are neglected.  We found that, depending by the value of the cutoff parameter, the presence of the background neutrino condensate can affect the formation and relaxation of wave dark matter halos. Nevertheless, for value of the cutoff of the order of a few eV, the two species can coexist showing only marginal differences with the only-wave dark matter case. These results open to the possibility of investigate about more complex cosmological scenarios involving the formation of dark matter halos.

\end{abstract}
\maketitle

\section{Introduction}

The majority of matter in the Universe seems to be Dark Matter (DM), an unseen component whose existence is inferred solely through gravitational effects, famously evidenced by galactic rotation curves \cite{Rubin1970,Rubin1982}. This non-baryonic constituent interacts minimally with Standard Model particles \cite{Bertone:2016nfn} and comprises approximately $84\%$ of the total matter density \cite{planck_collaboration_planck_2020}. Cosmological observations favor the Cold Dark Matter (CDM) model, which posits massive, non-relativistic, and collisionless particles. However, CDM faces significant inconsistencies at galactic scales. Key issues include the \textit{core-cusp} problem, and the \textit{missing satellite} problem, among the others. In the first case, $N$-body simulations predict dense, cuspy profiles ($\rho \sim r^{-1}$) for the DM halos \cite{Navarro:1995iw,de_Blok_CoreCusp}, contrasting with the cored profiles observed in dwarf galaxies \cite{deBlok2001,de_Blok_CoreCusp,Sharma2022}. Additionally, recent observational evidence indicates that the discrepancy between the cuspy profile and observed dark matter distributions persists across a wide range of galactic masses and Hubble types. The phenomenon is characterized not merely by the presence of a core, but by tight scaling relations between the core radius, the total halo mass, and the disk scale length—a relationship that standard CDM struggles to reproduce without significant baryonic feedback. Then, in the case of the {\em missing satellite} problem, simulations predict a smaller number of subhalos orbiting the Milky Way than are currently observed \cite{Read2019}. For extensive reviews on the subject, and the discussions about the need and emergence of new paradigms on observational grounds, we refer to \cite{Salucci2018, Salucci2019, DeMartino2020, Nesti2023,Salucci2025}. Moreover, among the numerous candidates for the identification of the nature and composition of DM proposed so far, ranging from weakly interacting massive particles to axions and sterile neutrinos, none has yet been confirmed experimentally \cite{AxionSterile2021,RediTesi2023,Yang2018}. 

Thus, in the search for a fundamental particle that resolves these small-scale challenges Fuzzy Dark Matter (FDM), also renamed Wave Dark Matter ($\psi$DM), emerges as a particularly intriguing candidate. $\psi$DM consists of ultralight bosonic particles (mass $m_\psi \sim 10^{-22}$ eV), whose large de Broglie wavelength ($\sim 1$ kpc) mandates the emergence of quantum behaviours on astrophysical scales \cite{hui_ultralight_2017,ferreira_ultra-light_2021,Eberhardt:2025caq}. Generated via a misalignment mechanism \cite{Preskill1983,Abbott1983}, $\psi$DM particles follow Bose-Einstein statistics \cite{Sikivie2009}, forming self-gravitating halos sustained by quantum pressure (due to the uncertainty principle). On cosmological time scale, $\psi$DM halos naturally develop a solitonic core, resulting in a cored mass density profile that successfully mitigates the core-cusp problem. Additionally, $\psi$DM imposes a minimum Jeans scale, suppressing the formation of low-mass structures and thereby providing a solution to the missing satellite problem \cite{Schive:2014dra,Schive:2014hza,Mocz:2017wlg, Davies2020}. 

Observationally, $\psi$DM has faced constraints that reveal a significant tension. Galactic-scale observations, such as velocity dispersion profiles in the bulge of Milky Way and in dwarf spheroidal (dSph) galaxies, favor a boson mass of $\sim 10^{-22}$ eV \cite{Chen2017,DeMartino2020PDU,Broadhurst2020, Pozo2020,Pozo2023,deMartino2023}. Conversely, cosmological constraints from the Lyman-$\alpha$ forest set a lower bound of $\gtrsim 2\times10^{-20}$ eV \cite{Armengaud2017,Rogers2021}, and the total mass of ultra-faint galaxies requires a mass $\gtrsim 10^{-21}$ eV \cite{Safarzadeh2020}.  This mass tension represents a critical open challenge for the FDM model, whose resolution could either confirm or rule out scenarios like the String Axiverse \cite{Pozo2023}. Therefore, distinguishing between $m_\varphi\sim 10^{-22}$~eV and $m_\varphi\sim 10^{-20}$~eV at the galactic scale is a crucial frontier for fundamental physics, warranting dedicated observational searches, such as pulsar timing with forthcoming facilities like the SKA \cite{DeMartino2017,Palencia:2025wjw}, strong gravitational lensing of highly magnified compact sources near radial arcs \cite{Palencia:2025wjw}, motion of stars around the supermassive black hole at the center of the Milky Way \cite{DellaMonica:2022kow,DellaMonica2023}, and proper motion measurements in next-generation astrometric satellites \cite{Furlanetto:2024qsb}.

All the previous results are based on a synergy between numerical simulations and observations, and pose their foundation on the assumption that $\psi$DM particles constitute the total amount of DM and  evolve as dictated by the gravitational potential well that, on turn, reflects the mass and energy density distribution of these particles. Nevertheless, as discussed by \cite{Burkert:2020laq}, standard wave dark matter models face challenges in matching these specific scaling relations, particularly the universal core column density. This broader observational tension provides a strong impetus for investigating modified frameworks, such as the wave dark matter-neutrino condensate interaction presented in this study, which may offer the dynamical flexibility required to satisfy these multi-scale observational constraints.

In the realm of Quantum Field Theory (QFT), the phenomenon of fermion mixing, as in the case of neutrinos of different flavors \cite{Alfinito:1995kx,Blasone:1995zc,Hannabuss:2000hy,Blasone:2002jv,Ji:2002tx}, gives rise to a vacuum condensate endowed with a non-zero energy density \cite{Capolupo:2006et,Capolupo:2012vf,Tarantino2012,Capolupo2018,Capolupo2020}.  
This flavor vacuum, analyzed in curved space time, possesses an equation of state analogous to that of cold DM, behaving effectively as a pressureless perfect fluid \cite{Capolupo2023,Carloni2022} and, therefore, it represents an additional gravitational source that might be  considered. In fact,  such a non-trivial vacuum energy cannot be removed through the normal-ordering procedure and, therefore, can contribute to the energy balance of the Universe \cite{Capolupo:2024ckb,Capolupo:2021cnt}.  
The formal properties of the condensate suggest that it could represent a subdominant component of DM, or at least a background field in which DM evolves \cite{Capolupo2020}.  
It is thus of great interest to identify analogous physical systems where similar effects can be studied under controlled conditions, as in the case of Rydberg atoms \cite{Capolupo:2019gmn}, and study experiments in which neutrino condensate effects can be detected \cite{Capolupo:2022hhr}. The analysis of neutrino capture rate on tritium, for non-relativistic neutrinos, can allow to
 prove the existence of quantum  condensation
effects \cite{Capolupo:2022hhr}. Experiments such as PTOLEMY \cite{Betti_2019}, which aim to detect the cosmic neutrino background, may provide indication of the existence of the dark-matter component arising from neutrino mixing. In this perspective, neutrinos deserve to be considered not only as active participants in cosmological evolution, but also as a potential background component shaping the dynamics of DM \cite{Carloni2022,Capolupo:2024ckb}.

Therefore, we want to focus our efforts on investigating the formation and relaxation of DM halos in the context of  $\psi$DM particles evolving on a background of neutrino vacuum condensate. To this aim, we will solve numerically the Schrödinger-Poisson system of equations to model the dynamical evolution of ultralight bosonic DM particles on cosmological time. In Section \ref{sec: psiDM}, we will introduce the $\psi$DM model pointing out its main features and the main results arisen from N-Body numerical simulations. Then, we will move in Section \ref{sec:simulations} to summarize the methodology used to integrate numerically the Schrödinger-Poisson system, and we will show, as a consistency check with the results in the literature, the outcomes of our simulations in the case of only-$\psi$DM particles. Then, in the Section \ref{sec:QFT}, we will introduce the theoretical framework of flavor field quantization in curved space-time with particular emphasis on the vacuum expectation value of the energy-momentum tensor, which will serve as an additional background source to the gravitational field. In the Section \ref{sec:psiDM+nu}, we will implement those results in the numerical methodology described in the Section \ref{sec:simulations}, and we will show the results in the Section \ref{sec:results}. Finally, we will give our main conclusion in the Section \ref{sec:conclusions}.

\section{The Wave Dark Matter}\label{sec: psiDM}

The non-relativistic dynamics of non-interacting $\psi$DM particles is governed by the coupled Schrödinger-Poisson (SP) system of equations. Therefore, the spatial density distribution of ultralight bosonic fields with masses around $m_\psi \simeq 10^{-22}\,\text{eV}$ is accurately described through the system \cite{Widrow:1993qq}:
\begin{equation}\label{eq:SP}
	\begin{aligned}
		i \frac{\partial}{\partial t} \psi(\vec{r}, t) &= 
		-\frac{\hbar^2}{2m_\psi} \nabla^2 \psi(\vec{r}, t) 
		+ m_\psi V(\vec{r}, t) \psi(\vec{r}, t), \\
		\nabla^2 V(\vec{r}, t) &= 4 \pi G T_{00},
	\end{aligned}
\end{equation}
where \(V(\vec{r}, t)\) denotes the gravitational potential, and the complex wavefunction \(\psi(\vec{r}, t)\) is normalized such that \(T_{00}=|\psi|^2\) represents the local mass density of the axionic DM component, \(\rho(\vec{r}, t)\).  Since all the particles in the halo share the same wavefunction, the probability density related to such wavefunction, \( |\psi|^2 \), traces the physical mass density of \( \psi \) DM \cite{Schive:2014dra, Mocz:2017wlg}. The SP system can be reformulated in a fluid-like representation through the Madelung transformation, allowing for an effective description of the \(\psi\)DM in terms of a fluid with density and velocity derived from the phase of the complex wavefunction \cite{Chavanis:2011zi,Suarez:2011yf}.
Since an analytic treatment of this coupled system is generally intractable, its evolution is best explored through numerical integration of Equation~\eqref{eq:SP} under suitable boundary and initial conditions.   

The N-body numerical simulations of the formation and evolution of $\psi$DM halos revealed the existence of a stable ground-state solution characterized by a solitonic core surrounded by wave-like fluctuations arising from interference patterns in the $\psi$DM wavefunction \cite{Woo:2008nn, Schive:2014dra, Mocz:2017wlg, Veltmaat:2018dfz}. The density profile of such a ground-state solution is well approximated by the following empirical formula \cite{Schive:2014dra,Schive:2014hza}:
\begin{equation}\label{eq:soliton}
	\rho_{\text{sol}}(r) = \rho_0 \left[ 1 + 9.1 \times 10^{-2} 
	\left( \frac{r}{r_c} \right)^2 \right]^{-8},
\end{equation}
where the central density is given by
\begin{equation}
	\rho_0 = 3.1 \times 10^6 
	\left( \frac{m_\psi}{2.5 \times 10^{-22}\,\text{eV}} \right)^{-2}
	\left( \frac{r_c}{\text{kpc}} \right)^{-4}
	\, \text{M}_{\odot}\,\text{kpc}^{-3}.
\end{equation}
Here, \(r_c\) denotes the solitonic core radius, defined as the radius at which the density drops to approximately half of its peak value. The cored configuration is due to an internal quantum pressure, which arises from the Heisenberg uncertainty principle over de Broglie wavelength scales, effectively counteracting gravitational collapse \cite{Hui:2021tkt}. Consequently, the density profile flattens in the central region. The soliton solution in Equation~\eqref{eq:SP} corresponds to the spherically symmetric, stationary solution of the SP equations, also known as a boson star or solitonic core. In  \cite{Schive:2014dra}, authors showed that such a solitonic core represents a general, virialized end state of $\psi$DM structures, largely independent of initial conditions. 

It is important to note that while the standard Schrödinger-Poisson framework, equation \eqref{eq:soliton}, successfully predicts cored central regions, these solitonic profiles often struggle to accurately reproduce the observed dark matter distributions across the full diversity of galactic scales. Recent studies \cite{Burkert:2020laq, DeLaurentis:2022nrv} have demonstrated that, while $\psi$DM can mimic certain scaling relations between core radii and virial masses, it fails to recover the observed universal core column density ($\Sigma_0 \approx 75M_\odot$ pc$^{-2}$). Instead, standard FDM models predict a steep dependence of the column density on the core radius, which is in complete disagreement with the constant $\Sigma_0$ found in nature. This suggests that if current $\psi$DM simulations hold, 'pure' wave dark matter may be ruled out as the sole origin of dark matter cores. This fundamental discrepancy strongly motivates the exploration of extended scenarios, such as the one presented in this work, where the inclusion of a neutrino condensate background introduces additional gravitational interactions that may modify the relaxation dynamics and the resulting core-halo relations to better align with astrophysical observations.

Since our aim is to perform three-dimensional numerical simulations of $\psi$DM halos in the presence of an additional background source to the gravitational field, in the following section we will describe how we carry out simulations in the simpler picture of the $\psi$DM, and we will show the results of simulations to ensure the effectiveness of the numerical methodology that will be employed later.

\subsection{Simulation of the Wave DM halo}\label{sec:simulations}

To  carry out the numerical simulations, the initial conditions \(\psi(\vec{r}, 0)\) are constructed as a superposition of \(n_c\) stationary solitonic profiles, each characterized by the same core radius \(r_c\), and randomly distributed within a cubic domain of side length \(L = 200\,\text{kpc}\).
The wavefunction is discretized on a uniform three-dimensional grid of resolution \(N^3 = 512^3\), ensuring a minimum of three grid points per solitonic core diameter. Time evolution is implemented via a second-order pseudo-spectral algorithm using a symplectic “kick-drift-kick” integration scheme, guaranteeing second-order temporal accuracy and spectral convergence in space (see, \textit{e.g.}, \cite{DellaMonica:2023vts, Mocz:2017wlg}
).  
The temporal update of the wavefunction is performed through the unitary evolution operator:
\begin{equation}
	\psi(\vec{r}, t + \Delta t) = e^{-i H \Delta t} \psi(\vec{r}, t),
\end{equation}
adopting Strang's splitting scheme \cite{Strang1968}, the Hamiltonian operator \(H = K + U\) (being \(K\) and \(U\) the kinetic and potential operators, respectively) is decomposed as:
\begin{equation}
	e^{-i H \Delta t} = e^{-i \frac{U \Delta t}{2}} 
	\, e^{-i K \Delta t} \, e^{-i \frac{U \Delta t}{2}}.
\end{equation}

The kinetic step is computed in Fourier space, where the operator \(K = \hbar^2 k^2 / (2m_\psi)\) acts on the transformed wavefunction \(\hat{\psi} = F[\psi]\).  
The potential step follows from the Poisson equation in Equation~\eqref{eq:SP}, and it is expressed in Fourier space as:
\begin{equation}
	\hat{V}(\vec{k}, t) = -\frac{4 \pi G \, \mathcal{F}(\rho - \bar{\rho})}
	{k^2},
    \label{eq:potential_step}
\end{equation}
where $\rho = |\psi|^2$ is the energy density and $\bar{\rho}$ denotes its mean value. The real-space potential is then recovered by an inverse Fourier transform:
\begin{equation}
	V(\vec{r}, t) = F^{-1}[\hat{V}(\vec{k}, t)].
\end{equation}

We implement the operators \( F \) and \( F^{-1} \) as 3D discrete fast Fourier transform (FFT) and inverse FFT \cite{Cooley:1965zz}. Following \cite{Schwabe:2016rze}, the time-step \( \Delta t \) satisfies the Courant–Friedrichs–Lewy condition and, therefore, the evolution of the wave-function in the SP system in Equation~\eqref{eq:SP} is made using an Hamiltonian operator that changes the phase of \( \psi \) by less than \( 2\pi \), and hence
 \begin{equation}
\Delta t \leq \max \left( \frac{m_\psi \, \Delta x^2}{6 \hbar}, \frac{\hbar }{m_\psi| V |_{\text{max}}} \right)\,.
 \end{equation}
Here  \( |V|_{\text{max}} \) is the maximum values of the gravitational potential on the grid whose spatial resolution is given by \( \Delta x \equiv \frac{L}{N} \). In such a way, we guarantee the stability and accuracy of the simulation. Finally, our integration scheme naturally imposes periodic boundary conditions. Since the total mass  \( M \)  is conserved, the halos that we simulate are constantly perturbed (reheated) by the reflecting waves at the box boundary. Therefore, the final state never relaxes to a perfect solitonic equilibrium configuration as in \cite{Schwabe:2016rze} which is analytically expressed in Equation~\eqref{eq:soliton} but, similarly to \cite{Schive:2014dra,Mocz:2017wlg}, is closer enough to the cosmological scenario in which waves coming from distant halos arrive from all directions. 
 
The initial total mass \( M \) of the halo is given once  the boson mass \( m_\psi \), the number \( n_c \), and the radii \( \{r_{c,i}\}_{i=0}^{n_c} \) of the initial solitons are set. Moreover, the total mass  is directly computed from the solitonic mass density profile in Equation~\eqref{eq:soliton} and results to be (for each soliton)
\begin{equation}
	M_s(r_{c,i}) \approx 2.2 \times 10^{10} \left( \frac{m_\psi}{10^{-23} \, \text{eV}} \right)^{-2} \left( \frac{r_{c,i}}{\text{kpc}} \right)^{-1} \, M_\odot,
\end{equation}
which leads to the following total (conserved) mass of our simulations
\begin{equation}
	M = \sum_{i=0}^{n_c} M_s(r_{c,i}).
\end{equation}

\begin{figure*}[ht!]
	\includegraphics[width=2\columnwidth]{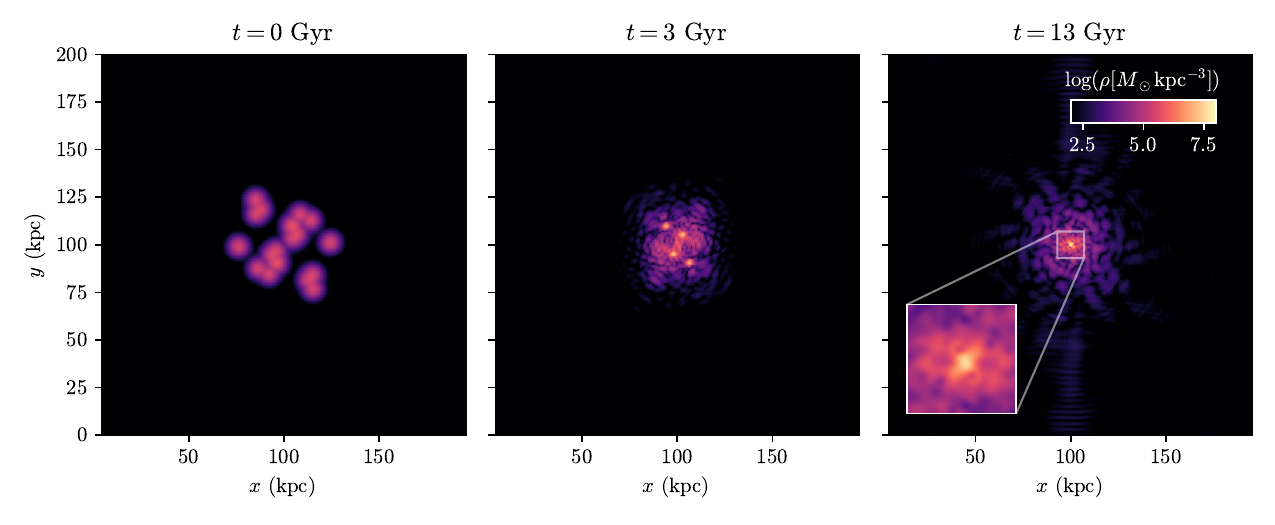}
	\caption{The figure shows a specific but representative configuration of the initial conditions: \( n_c = 20 \) and  \( M = 5\times10^8 M_\odot \). The left, middle, and right snapshots show equatorial slices of the merger evolution at  0.0, 3.0 and 13 Gyr, respectively. In the right panel, the inset shows the innermost region of the virialized halo with the solitonic core. }\label{fig:fig1}
\end{figure*}

We run a total of  100 simulations with different initial conditions. In each simulation, we have fixed the total mass to \( M \sim 5\times10^8 M_\odot \), and the boson mass to \( m_\psi = 2.1 \times 10^{-22} \, \text{eV} \). Then, we have generated randomly different sets of values for  \( n_c \) and \( \{r_{c,i}\}_{i=0}^{n_c} \) in such way that the total mass is conserved, and random central positions for the solitons within the simulation box, ensuring that the mass centroid is always fixed at the center of the box. This allows us to verify that the final configuration of each simulation is self-similar, independently of the initial conditions. To show the effectiveness of our procedure, we show in Figure \ref{fig:fig1} a specific but representative configuration of the initial conditions with \( n_c = 20 \) and  \( M = 5\times10^8 M_\odot \). The left, middle, and right snapshots show equatorial slices of the merger evolution at  0.0, 3.0 and 13 Gyr, respectively. In the right panel, the inset shows the innermost region of the virialized halo with the solitonic core. 

\begin{figure}[ht!]
	\includegraphics[width=\columnwidth]{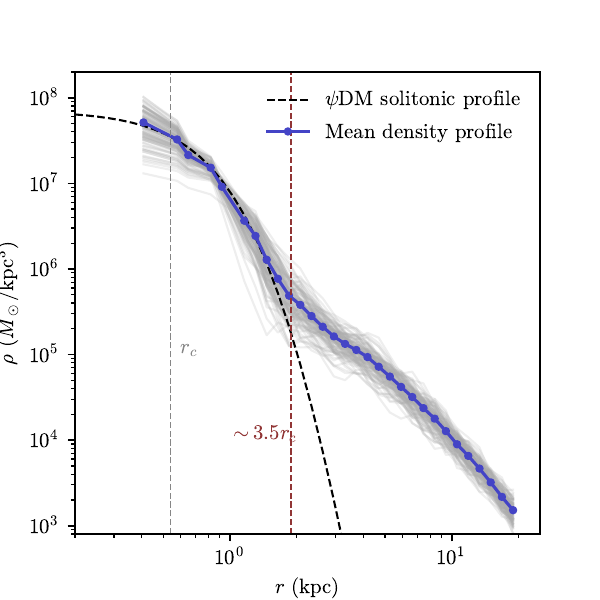}
	\caption{The figure depicts the radial mean mass density profile averaged over the simulations (blue line), and also shows the radial mass density profiles emerging from each of the 100 simulations (light-grey lines). We overplot the solitonic mass density profile in Equation~\eqref{eq:soliton} with a central density of \( \rho_0 \sim 7 \times 10^7 M_\odot/\text{kpc}^3 \) and a solitonic core radius of  \( r_c = 550 \, \text{pc} \), as the black dashed line. Starting from the break in the inner slope occurring at \( \sim 3.5 r_c \), the outermost slope follows a NFW-like profile emerging when making an azimuthal average of the quantum fluctuations due to the interference pattern of the wavefunction.}\label{fig:fig2}
\end{figure}

Finally, in Figure \ref{fig:fig2}, we depict the radial mean mass density profile averaged over the simulations (blue line), and we also show the radial mass density profiles emerging from each simulations (light-grey lines).   The innermost part is well-described by  the solitonic mass density profile in Equation~\eqref{eq:soliton} with a central density of \( \rho_0 \sim 7 \times 10^7 M_\odot/\text{kpc}^3 \) and a solitonic core radius of  \( r_c = 550 \, \text{pc} \), which is overplotted as the black dashed line. The outer part of the radial mass density profile,  starting from the break in the inner slope occurring at \( \sim 3.5 r_c \), follows a NFW-like profile ($\rho\propto r^{-3}$) emerging when making an azimuthal average of the quantum fluctuations due to the interference pattern of the wavefunction that dominate the outermost region of the halo.

After having demonstrated the effectiveness of the methodology in simulating the evolution of ultralight bosons, we want to investigate the formation and relaxation of $\psi$DM halos when ultralight bosons evolve on a background of neutrinos condensate. To this aim, in the next section, we will summarize the theoretical framework underlying the quantization of flavor fields in curved backgrounds which leads to the formation of neutrino vacuum condensate \cite{Capolupo:2020wlx,Capolupo:2024ckb,Capolupo:2021cnt,Capolupo:2023wri} .

\section{Overview of Flavor Field Quantization in Curved Spacetime}\label{sec:QFT}

 Let us consider the action for free neutrino
   Dirac fields, $\nu_1(x)$, $\nu_2(x)$, and $\nu_3(x)$,   with masses $M_1$, $M_2$, and $M_3$,  
\begin{equation}
	S = \int d^4x \, \sqrt{-g} \left[ 
	\frac{i}{2} \sum_{i=1}^{3} 
	\left( \bar{\nu}_i \gamma^{\mu} D_{\mu} \nu_i 
	- D_{\mu} \bar{\nu}_i \gamma^{\mu} \nu_i \right) 
	- M_i \bar{\nu}_i \nu_i 
	\right]\,.
	\label{eq:action}
\end{equation}
Here $g_{\mu\nu}$ denotes the spacetime metric with determinant $g$, and $\bar{\nu}_i = \nu_i^{\dagger} \gamma^0$ is the adjoint field. All quantities in this section will be expressed, for sake of simplicity, in Plank units, {\em i.e.} $G=c=\hbar=1$. Introducing the tetrad basis $e^a_{\ \mu}$, the curved-space gamma matrices and the spinorial covariant derivative take the form 
$\gamma^{\mu} = e^{\mu}_{\ a} \gamma^{a}$ and $D_{\mu} = \partial_{\mu} + \Gamma_{\mu}$, respectively, where the spin connection is given by 
$\Gamma_{\mu} = \tfrac{1}{8}\,\omega_{\mu ab} [\gamma^{a}, \gamma^{b}]$, 
and $\omega_{\mu ab}$ are the spin connection coefficients determined by the tetrad field. 
 
Let $\{ U_{i,\lambda}, V_{i,\lambda} \}$ denote a complete orthonormal set of solutions of the curved-space Dirac equations $	(i \gamma^{\mu} D_{\mu} - M_i)\, \nu_i = 0,$
where $\lambda$ labels momentum and spin degrees of freedom.  Each massive field can then be decomposed as follows:
\begin{equation}
	\nu_i(x) = 
	\sum_{\lambda} \int d^3 q \,
	\big[ 
	a_{i,\lambda}(\mathbf{q}) \, U_{i,\lambda}(\mathbf{q},x)
	+ b_{i,\lambda}^{\dagger}(\mathbf{q}) \, V_{i,\lambda}(\mathbf{q},x)
	\big],
	\label{eq:field_expansion}
\end{equation}
where $a_{i,\lambda}$ and $b_{i,\lambda}$ are the annihilation operators for particles and antiparticles, respectively.   The vacuum state associated with the mass fields, denoted $|0\rangle$, is defined as $a_{i,\lambda} |0\rangle = 0 = b_{i,\lambda} |0\rangle$ for any $i$ and $\lambda$. We denote with $J_{\Theta}(\tau)$ the generator of mixing transformations, defined as:
\begin{widetext}
\begin{equation}\label{eq: neutrinoflavour}
	\begin{pmatrix}
		\nu_e(x) \\[4pt]
		\nu_\mu(x) \\[4pt]
		\nu_\tau(x)
	\end{pmatrix}
	=
	J^{-1}_{\boldsymbol{\Theta}}(x)
	\begin{pmatrix}
		\nu_1(x) \\[4pt]
		\nu_2(x) \\[4pt]
		\nu_3(x)
	\end{pmatrix}
	J_{\boldsymbol{\Theta}}(x)
	=\\
	\begin{pmatrix}
		c_{12} c_{13} & s_{12} c_{13} & s_{13} e^{-i\delta} \\
		-s_{12} c_{23} - c_{12} s_{23} s_{13} e^{i\delta} & c_{12} c_{23} - s_{12} s_{23} s_{13} e^{i\delta} & s_{23} c_{13} \\
		s_{12} s_{23} - c_{12} c_{23} s_{13} e^{i\delta} & -c_{12} s_{23} - s_{12} c_{23} s_{13} e^{i\delta} & c_{23} c_{13}
	\end{pmatrix}
	\begin{pmatrix}
		\nu_1(x) \\[4pt]
		\nu_2(x) \\[4pt]
		\nu_3(x)
	\end{pmatrix}.
\end{equation}
 
where $\delta$ is the Dirac CP-violating phase,   
$s_{ij} = \sin\theta_{ij}$ and $c_{ij} = \cos\theta_{ij}$, and $\theta_{ij}$ with $i,j=1,2,3$ are the mixing angles.
The flavour annihilators  are given by
\begin{align}
	a_{\mathbf{q},\sigma}(\tau) 
	&= J^{-1}_{\boldsymbol{\Theta}}(\tau)\, a_{\mathbf{q},i}\, J_{\boldsymbol{\Theta}}(\tau) \nonumber\\
	&= c_{12} c_{13}\, a_{\mathbf{q},1}
	+ s_{12} c_{13} 
	\sum_{\lambda'} 
	\left( 
	\Xi^{12}_{\mathbf{q},\lambda,\lambda'}\, a_{\mathbf{q},2,\lambda'}
	+ \Omega^{12}_{\mathbf{q},\lambda,\lambda'}\, b^{\dagger}_{\mathbf{q},2,\lambda'} 
	\right)
	+ e^{-i\delta} s_{13} 
	\sum_{\lambda'} 
	\left( 
	\Xi^{13}_{\mathbf{q},\lambda,\lambda'}\, a_{\mathbf{q},3,\lambda'}
	+ \Omega^{13}_{\mathbf{q},\lambda,\lambda'}\, b^{\dagger}_{\mathbf{q},3,\lambda'}
	\right),
	\label{eq:flavor_annihilator}
\end{align}
 \end{widetext}
analogous expressions hold for the remaining flavour operators.
The Bogoliubov coefficients appearing in the above equation   are defined as $	\Xi^{L,M}_{\mathbf{q},\lambda,\lambda'}(\tau) 
	= (U_{L,\mathbf{q},\lambda}, U_{M,\mathbf{q},\lambda'})_\tau,$ $
	\Omega^{L,M}_{\mathbf{q},\lambda,\lambda'}(\tau)
	= (U_{L,\mathbf{q},\lambda}, V_{M,\mathbf{q},\lambda'})_\tau,$
where $(\cdot,\cdot)_\tau$ is the Dirac inner product defined on the hypersurface $\Sigma_\tau$.  
The operators  in Equation~\eqref{eq:flavor_annihilator} annihilate the time-dependent flavour vacuum, $|0(\tau)\rangle_f = J^{-1}_{\boldsymbol{\Theta}}(\tau)\, |0\rangle, $
which belongs to a Fock space unitarily inequivalent to that of the mass eigenstates.  
$|0(\tau)\rangle_f$ exhibits a   structure of  a condensate of particle–antiparticle pairs with definite masses. 
As a consequence, its energy–momentum content is non-vanishing.
 The energy–momentum tensor derived from the action in Equation~\eqref{eq:action} reads:
\begin{equation}
	T^{(3)}_{\mu\nu}(x) = 
	\frac{i}{2} \sum_{L=1}^{3} 
	\left[
	\bar{\nu}_L \gamma_{(\mu} D_{\nu)} \nu_L 
	- D_{(\mu} \bar{\nu}_L \gamma_{\nu)} \nu_L
	\right].
	\label{eq:stress_tensor}
\end{equation}

 The instantaneous energy–momentum density of the flavor vacuum is  
\begin{equation}
	\mathcal{T}^{(3)}_{\mu\nu}(x) = 
	\langle 0(\tau) | : T^{(3)}_{\mu\nu}(x) : | 0(\tau) \rangle,
	\label{eq:vev_stress_tensor}
\end{equation}
where    $:..:$ denotes the normal ordering  with respect to the mass vacuum \cite{Capolupo:2021cnt}.  We note that the neutrino condensate originates from particle mixing and arises as a consequence of the non-unitary inequivalence between the Fock spaces associated with the mass and flavor representations of neutrinos \cite{Blasone:1995zc,Capolupo:2006et,Capolupo:2016pbg}. Consequently, the existence of this condensate is independent of the spacetime metric, although its specific configuration depends on the geometry considered. In the following, we restrict our analysis to a static, spherically symmetric spacetime, which is particularly relevant for astrophysical systems and the description of galaxies.

\subsection{Vacuum Expectation Value of the Energy--Momentum Tensor}

In isotropic coordinates, the static and spherically symmetric  spacetime metric takes the form

\cite{Egorov:2024eas}: 
\begin{equation}\label{eq:metric}
	ds^2 = f(r)\, dt^2 - g(r)\, (dx^2 + dy^2 + dz^2),
\end{equation}
 where $r = \sqrt{x^2 + y^2 + z^2}$, and the functions $f(r)$ and $g(r)$ are assumed to be sufficiently smooth and regular over the region of interest. According to the analysis developed in \cite{Capolupo:2024ckb}, all purely spatial components of both $\mathcal{T}^{(3)}_{\mu\nu}(x)$ vanish identically,
\begin{equation}
	\mathcal{T}^{(3)}_{jk}(x) = 0, \qquad \forall j,k = 1,2,3.
\end{equation}

Furthermore, the remaining off-diagonal components $\mathcal{T}^{(3)}_{0i}$ are found to vanish in a broad class of physically relevant geometries, such as Friedmann–Lemaître–Robertson–Walker (FLRW) spacetimes and static spherically symmetric metrics \cite{Capolupo:2021cnt}.   Consequently, the only non-zero contribution arises from the temporal component $\mathcal{T}^{(3)}_{00}$.

Under these conditions, the energy–momentum tensor associated with the flavor vacuum assumes the form of a pressureless perfect fluid, with   equation of state of dust or cold  dark matter: $w=  p /\rho = 0$.  
Where $\rho = \mathcal{T}^{(3)}_{00}$ is the energy density $p=\mathcal{T}^{(3)}_{jj} =0$  is the pressure of the flavor vacuum condensate.  
The flavor vacuum energy content $\mathcal{T}^{(3)}_{00}$ can be expressed as 
 \begin{equation}
 	\mathcal{T}^{(3)}_{00} = 4\, \mathcal{K}^{(3)}(1+4V(r)),    
 \end{equation}
where $\mathcal{K}^{(3)}$ is the contribution of  flavour vacuum condensate in the Minkowski metric \cite{Capolupo:2024ckb}. This quantity is ultraviolet divergent. We therefore introduce a momentum cutoff $\Lambda$ to regularize the integral defining the condensate energy density (for more details we refer to \cite{Capolupo:2024ckb}). With this regularization, $\mathcal{K}^{(3)}$ takes the explicit form: 
\begin{widetext}
\begin{equation}
	\begin{aligned}
		4\, \mathcal{K}^{(3)} 
		= \frac{1}{2\pi^2} \Bigg\{ &
		\Big[
		\Lambda \sqrt{\Lambda^2 + M_1^2}
		- M_1^2 \ln\!\left(
		\frac{\Lambda + \sqrt{\Lambda^2 + M_1^2}}{M_1}
		\right)
		\Big]
		\Big[
		M_1^2 
		- M_1 M_2 s_{12}^2 c_{12}^2 c_{13}^2
		- M_1 M_2 | s_{12} c_{23} - c_{12} s_{23} s_{13} e^{i\delta} |^2
		\Big] \\
		& - M_1 M_3 | s_{12} s_{23} + c_{12} c_{23} s_{13} e^{i\delta} |^2
		- M_2 M_3 | c_{12} s_{23} - s_{12} c_{23} s_{13} e^{i\delta} |^2 \\
		& + \Big[
		\Lambda \sqrt{\Lambda^2 + M_2^2}
		- M_2^2 \ln\!\left(
		\frac{\Lambda + \sqrt{\Lambda^2 + M_2^2}}{M_2}
		\right)
		\Big]
		\Big[
		M_2^2 
		- M_1 M_2 s_{12}^2 c_{13}^2
		- M_2 M_3 | c_{12} c_{23} + s_{12} s_{23} s_{13} e^{i\delta} |^2
		\Big] \\
		& - M_2 M_3 | c_{12} s_{23} + s_{12} c_{23} s_{13} e^{i\delta} |^2 \\
		& + \Big[
		\Lambda \sqrt{\Lambda^2 + M_3^2}
		- M_3^2 \ln\!\left(
		\frac{\Lambda + \sqrt{\Lambda^2 + M_3^2}}{M_3}
		\right)
		\Big]
		\Big[
		M_3^2 
		- M_3^2 c_{23}^2 c_{13}^2
		- M_1 M_3 s_{12}^2 s_{13}^2
		- M_2 M_3 s_{23}^2 c_{13}^2
		\Big]
		\Bigg\}.
		\label{eq:K3}
	\end{aligned}
\end{equation}
\end{widetext}
 Notice  that $\mathcal{K}^{(3)}$ vanishes in the absence of mixing, \emph{i.e.}, when all the mixing angles $\theta_{AB}$ are equal to zero.  We consider the weak-field regime of the metric  in Equation~\eqref{eq:metric}, by setting
\begin{equation}
	f(r) = 1 + 2V(r), 
	\qquad 
	g(r) = 1 - 2V(r).
\end{equation}
Here the gravitational potential $V(r)$ is treated as a small perturbation.  
Retaining only terms up to first order in $V(r)$, and assuming that the flavor vacuum is the only source of the energy momentum,  the Poisson equation due to $V(r)$ is  
\begin{equation}
	\nabla^2 V(r) = 4\pi G\, \mathcal{T}^{(3)}_{00}.
	\label{eq:poisson_weakfield}
\end{equation}

\section{Axion Field Dynamics in a Neutrino Condensate Background}\label{sec:psiDM+nu}

The coupling between light axions and fermionic degrees of freedom constitutes a central feature in several extensions of the Standard Model, particularly within the framework of axion-like particle phenomenology (see, \textit{e.g.}, \cite{Dolan:2014ska,Carmona:2021seb,DallaValleGarcia:2023xhh}). Originally introduced to resolve the strong CP problem in quantum chromodynamics \cite{Peccei:1977hh,PhysRevD.101.083014,Raffelt:1996wa,Raffelt:2006cw,Capolupo:2019peg,DeMartino:2017qsa,Capolupo:2019xyd}, axions can interact with fermions through pseudoscalar or derivative couplings.   In the present analysis, however, our attention is not directed toward such interactions but rather focused on the dynamical evolution of the axion field in the presence of a neutrino vacuum condensate background.  In this scenario, the axion field evolves under the influence of both its self-gravity and the additional gravitational field generated by the neutrino vacuum condensate, while couplings between $\psi$DM and neutrinos, and the dynamical evolution of the neutrino vacuum background are neglected. Therefore, the Poisson equation  must be recast accordingly as
\begin{equation}\label{eq:Poisson_nu}
		\nabla^2 V(\vec{r}, t) = 4 \pi G \biggl[ |\psi(\vec{r}, t)|^2 + \mathcal{T}^{(3)}_{00}\biggl]\,,
\end{equation}
while the Schrödinger equation remains unaltered. To carry out numerical simulations with the modified Poisson equation, the potential operator in Equation~\eqref{eq:potential_step} that follows from the Poisson equation in Equation~\eqref{eq:Poisson_nu}, can be expressed in Fourier space as
\begin{equation}
	\hat{V}(\vec{k}, t) = -\frac{4 \pi G \, \mathcal{F}(\rho - \bar{\rho}) + 16 \pi K^{(3)} \beta}
	{k^2 + \frac{64 \pi K^{(3)} \beta}{c^2}},
\end{equation}
where  $\beta \equiv G / (\hbar^3 c^5)$.  In this formulation, we are thus neglecting at first order the free energy of the neutrino component.

Finally, we follow the same procedure explained in Section \ref{sec:simulations}.  We run a total of  100 simulations with the same initial conditions used for the $\psi$DM simulations. In each simulation, the total mass is set to \( M \sim 5\times10^8 M_\odot \), and the boson mass is fixed to \( m_\psi = 2.1 \times 10^{-22} \, \text{eV} \). In such a way, we will be able to discuss possible differences with the only-$\psi$DM simulations arising by the introduction of the neutrino backgrounds. Moreover, in each simulations we have fixed the parameters 
$s_{ij} = \sin\theta_{ij}$ and $c_{ij} = \cos\theta_{ij}$  used for the three mixing angles 
$\boldsymbol{\Theta} = (\Theta_{12}, \Theta_{13}, \Theta_{23})$ to: 
$s_{12}=0.0307$, $s_{13}=0.0219$, $s_{23}=0.558$,  $c_{ij}=\sqrt{1-s_{ij}^2}$. Then, the neutrino mass eigenstates with masses $M_1$, $M_2$, and $M_3$ have been fixed to 1 meV, $8.73$ meV and $50.3$ meV, respectively. On the contrary, the value of the ultraviolet momentum cutoff $\Lambda$ is varied to study its impact.
 Thus, we run several sets of simulations for the following values of $\Lambda=[1, \,5, \, 50,\,100, \,500]$ eV. We considered small values of the cutoff $\Lambda$, since for $\Lambda \sim \mathcal{O}(1\,\mathrm{eV})$, the flavor vacuum energy density assumes values compatible with the average galactic dark matter density \cite{Capolupo2023,Carloni2022}. 
 The introduction of such a cutoff is required 
because the integral defining the condensate energy density is ultraviolet divergent.
From a mathematical point of view, this divergence originates from the contribution of high momentum modes, which dominate the integral when the 
momentum is extended to arbitrarily large values. However, the physical properties of the neutrino mixing condensate are determined by the momentum 
dependence of the Bogoliubov coefficients that characterize the flavor vacuum.

These coefficients exhibit a nontrivial momentum dependence and attain their maximum for momenta of order
$
k \sim \sqrt{m_1 m_2},
$
which defines the characteristic momentum scale associated with neutrino 
mixing \cite{Capolupo:2024ckb}. For realistic neutrino masses this scale lies in the sub-eV region, indicating that the nontrivial structure of the condensate is mainly generated by modes with momenta comparable to the neutrino mass scale.

Moreover, quantum field–theoretical effects become progressively suppressed at high energies. In particular, the Bogoliubov coefficients vanish in the large momentum limit \cite{Capolupo:2024ckb}.  Choosing $\Lambda$ in the eV range therefore reflects the intrinsic scale set by the 
neutrino masses and leads to vacuum energy densities consistent with  
galactic dark matter density.

\section{Results and Discussion}\label{sec:results}

The simulation successfully models the dynamical evolution of wave DM halos over a background of neutrino condensate. The evolution is governed by the SP system of equations \eqref{eq:SP} where the Poisson equation of wave DM is replaced with the one that includes also the energy density of the neutrino vacuum condensate, as reported in Equation~\eqref{eq:Poisson_nu}. While in our baseline model (\textit{i.e.}, $\psi$DM), that will serve as comparison, the gravitational field is determined by the energy density distribution of the ultralight bosons, in this model we add another component that contributes to the gravitational potential through its vacuum energy density. Although the dynamical evolution of the neutrino component, as well as its coupling to the ultralight boson field are neglected, we found that the neutrino vacuum condensate can alter the relaxation of halos with respect to the baseline model. In fact, the emergence of solitonic cores and wave interference patterns are observed, consistent with theoretical predictions of wave DM, but their formation and emergence depends on the value of the ultraviolet momentum cutoff parameter $\Lambda$. 

\begin{figure*}[!ht]
    \centering
    \includegraphics[width=\textwidth]{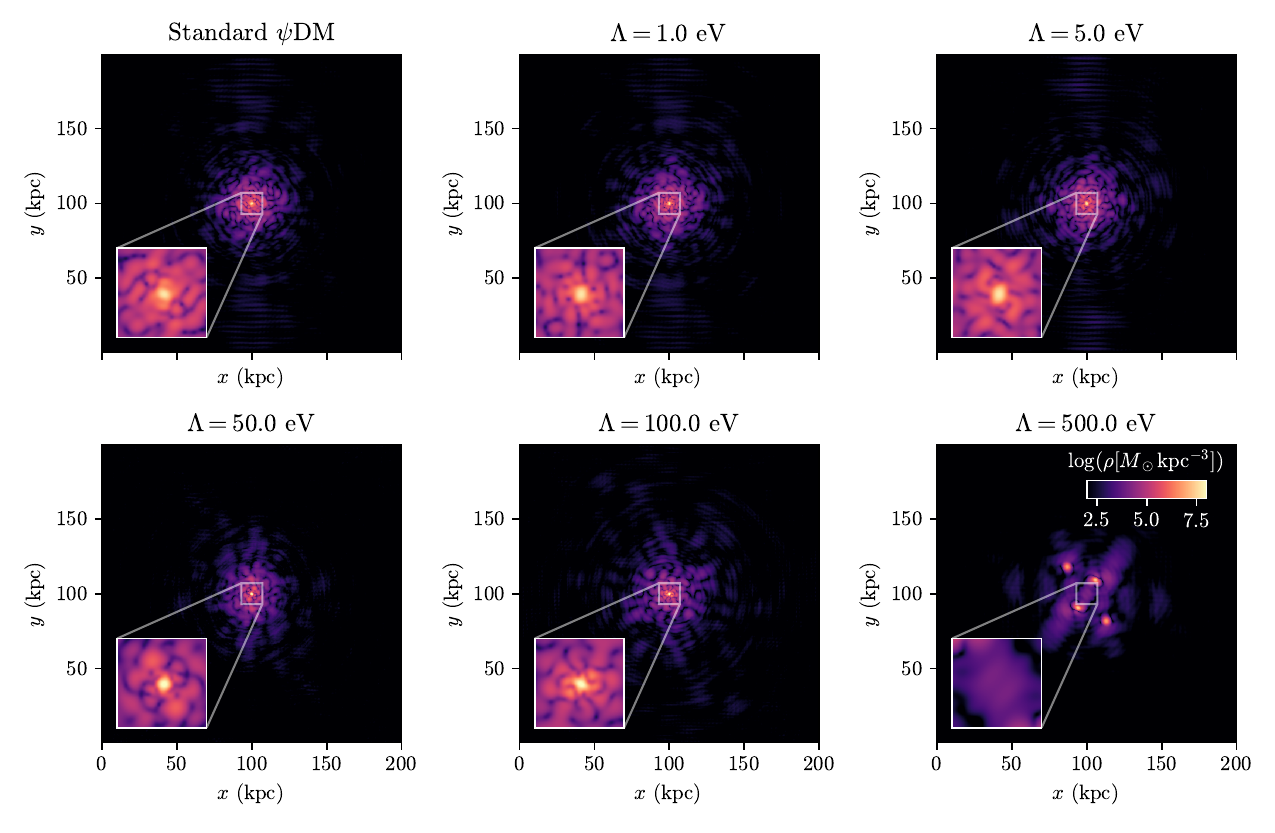}
    \caption{The figure illustrates the final state after 13 Gyrs of evolution of a specific realization of the initial conditions. The equatorial snapshots show the final state of the evolution of  only $\psi$DM and in the case of neutrino-$\psi$DM simulations with different values of the cutoff parameters (as reported on top of each panel), for comparison.  The insets show the innermost region of the halo with the solitonic core that emerges in all simulations except for the highest value of $\Lambda $ taken into consideration. All plots use the same color scale for the density, as reported in the last plot.}
    \label{fig:simulations_final_results}
\end{figure*}

\begin{figure}[!ht]
    \centering
    \includegraphics[width=\columnwidth]{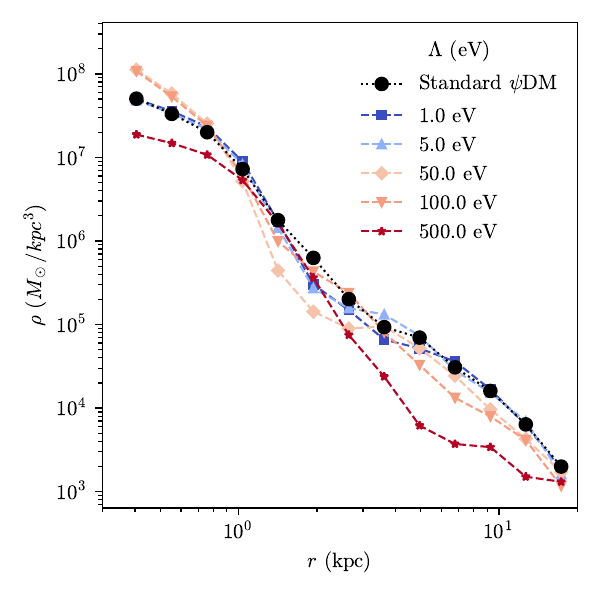}
    \caption{The figure depicts the mass density profile of the halos corresponding to different values of $\Lambda$. As black circular point, we depicted the results from the $\psi$DM simulation, while as blue squares, light blue triangles, orange diamonds, coral nablas and red stars, we depicted the results of the new simulations for the values of $\Lambda$ equal to 1, 5, 50 ,100 and 500 eV, respectively. The main differences for the highest value of the cutoff appear in the central value of the mass density and at the transition point.}
    \label{fig:simulations_final_results_profiles}
\end{figure}

In Figure \ref{fig:simulations_final_results}, we illustrate the final state at 13 Gyrs of the evolution of a given realization of the initial conditions, for $\psi$DM-only and in the case of neutrino-$\psi$DM simulations with different values of the ultraviolet momentum cutoff parameter (as reported on top of each panel), for comparison.  The insets show the innermost region of the halo with the solitonic core that emerges in all simulations except for the highest value of $\Lambda $ taken into consideration (\textit{i.e.} 500 eV). In fact, we found qualitatively that increased values of the ultraviolet momentum cutoff do not allow for the formation of the solitonic core, nor the emergence of the quantum fluctuations that characterize the outermost part of the $\psi$DM-only halo. In Figure \ref{fig:simulations_final_results_profiles}, we depicted the mass density profile of the halos  corresponding to different values of $\Lambda$ to be compared with our baseline model. With black circular points, we depicted the results from the $\psi$DM simulation, while as blue squares, light blue triangles, orange diamonds, coral nablas and red stars, we depicted the results of the new simulations for the values of $\Lambda$ equal to 1, 5, 50, 100 and 500 eV, respectively. {Neglecting the expected random fluctuations, we found differences appearing at the transition point from the solitonic cores to the outer profile. Stronger differences are found for higher values of $\Lambda$. In fact, we found that the central density appears to be higher for greater values of $\Lambda$ and the transition point to the outer mass density profile appears to fall at inner radii compared to the baseline model, signaling an increase in compactness of the soliton (and, hence, a more cuspy mass density profile) as the greater values of the cutoff parameter are considered. This behavior breaks going from $\Lambda = 100$ eV to $\Lambda = 500$ eV, as visually shown in the inset plot of Figure \ref{fig:simulations_final_results}. In this case the increased gravitational potential from the neutrino condensate prevents the halo from reaching a fully relaxed and virialized configuration, leading instead to the formation of several compact sub-halos. Therefore, we argue that in order to make these two dark components coexist and form virialized DM halos within the Hubble time, the highest allowed value of the ultraviolet momentum cutoff is of the order of 100 eV, though still showing some important differences with the standard $\psi$DM halo as greater values of $\Lambda$ are considered.}

\begin{figure}[!ht]
    \centering
    \includegraphics[width=0.95\columnwidth]{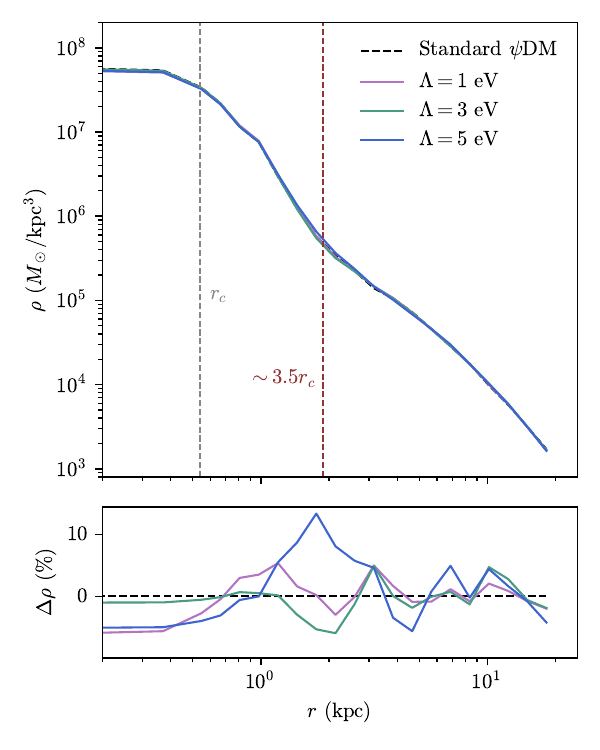}
    \caption{The figure depicts the final halo mass density profiles for the standrad $\psi$DM (dashed black line), and for $\Lambda= 1$ eV (magenta line), $\Lambda= 3$ eV (green line),  and $\Lambda= 5$ eV (blue line), respectively. All profile show an inner solitonic core and an outer NFW-like mass density distribution. In the lower panel, we show the relative difference with the  $\psi$DM-only mass density profile which oscillates between a few percent up to about 13\% due to the different radius of the transition between the two regions.}
    \label{fig:averaged-profiles-neutrino}
\end{figure}

Finally, we choose a few representative values of $\Lambda$ to run the full set of simulations (over the full array of 100 different initial conditions) and study the final DM density profile to check whether or not they show differences with the $\psi$DM-only simulations by averaging-out random fluctuations. In Figure \ref{fig:averaged-profiles-neutrino}, we show the final halo mass density profiles averaged over the 100 simulations for $\Lambda= 1$ eV (magenta line), $\Lambda= 3$ eV (green line), and $\Lambda= 5$ eV (blue line), respectively. Qualitatively, all profiles show similar features: an inner solitonic core and an outer NFW-like mass density distribution. The main quantitative differences with the $\psi$DM-only simulations appear at the transition point between the two regions. In the lower panel, we show the relative difference with the  $\psi$DM-only mass density profile. These differences oscillate between a few percent up to about 13\%,  and they are due to the different radius of the transition between the two regions. The change in the radius of the transition is the main difference of this scenario with respect to  the $\psi$DM-only one. In fact, even with the additional gravitational source of neutrino vacuum condensate, the halo is virialized. Therefore, we argue that, at least for small values of the ultraviolet momentum cutoff parameter of the order of a few eV, the two species can coexist and form a DM halo. 

\section{Conclusion}\label{sec:conclusions}
This work presents a numerical analysis of the Schrödinger-Poisson system to study the impact of the existence of a neutrino vacuum condensate, acting as an additional background source of gravitational field, on the formation and relaxation of halos composed of ultralight $\psi$DM. We have first validated our methodology, ensuring that it effectively captures key features of $\psi$DM-only dynamics, such as the formation of distinctive solitonic cores and quantum fluctuation patterns that solely depend on the mass of the ultralight boson considered. Then, we have added the neutrino condensate component and performed several simulations to asses how the final state of the simulation after one Hubble time is affected by the presence of the extra component, as a function of the ultraviolet momentum cutoff parameter. Our implementation is based on two key assumptions: first, the neutrinos vacuum condensate  just acts as an additional gravitational source without evolving due to the interaction with the ultralight bosons constituting the $\psi$DM particles, and/or  its own gravitational field. And, second, self-interactions and couplings between the two species are also neglected. Under these assumption we have found the the two species can coexist and form a relaxed halo within the Hubble time if the value of the ultraviolet momentum cutoff is lower than a few hundreds of eV. In fact, setting $\Lambda$ to 500 eV or higher would prevent the formation of a virialized halo. Finally, the main physical effect is that the solitonic core that forms in the $\psi$DM model, is now more compact with an higher central density and smaller core radius. Future analyses will focus on removing the assumptions to make more comprehensive (and computationally expensive) numerical simulations allowing the neutrino vacuum condensate to evolve jointly with the ultralight bosons and, eventually, form the final halo to point out some astrophysical signature of this alternative picture.

\section*{Acknowledgements} 
AC and SM acknowledge partial financial support from MUR and Istituto
Nazionale di Fisica Nucleare (INFN), Sezione di Napoli and Gruppo Collegato di Salerno, Iniziativa Specifiche QGSKY. AC also acknowledges the COST Action CA1511 Cosmology and Astrophysics Network for Theoretical Advances and Training Actions (CANTATA);
IDM acknowledges support from the grant PID2024-158938NB-I00 (Spanish Ministerio de Ciencia e Innovación and FEDER “A way of making Europe”). IDM also acknowledges support under the project SA097P24 funded by Junta de Castilla y Le\'on. RDM acknowledges financial support provided by FCT – Fundação para a Ciência e a Tecnologia – through the ERC-Portugal program Project “GravNewFields”, and through grant No. \href{https://doi.org/10.54499/UID/PRR/00099/2025}{UID/PRR/00099/2025} and grant No. \href{https://doi.org/10.54499/UID/00099/2025}{UID/00099/2025} to the Center for Astrophysics and Gravitation (CENTRA/IST/ULisboa). Finally, we acknowledge the center of Supercomputación Castilla y León (SCAYLE) for providing supercomputing facilities.

\bibliographystyle{apsrev4-1}
\bibliography{INSPIRE.bib}

\end{document}